\documentclass[amssymb,aps,twocolumn,floats,showpacs]{revtex4-1}
\usepackage{color,graphicx,pstricks,float}
\usepackage{natbib}

\usepackage{fancybox}
\usepackage{epsfig}
\usepackage{graphicx}
\usepackage{tabularx}
\usepackage{inputenc}
\usepackage{amsmath}
\usepackage{amssymb}
\usepackage{gensymb}
\usepackage{array,xcolor}
\usepackage{amsbsy}

\usepackage{color,graphicx,pstricks,float}

\usepackage{url,hyperref}
\hypersetup{
	colorlinks=true,
	linkcolor = blue,
	filecolor = blue,      
	urlcolor = blue,
	citecolor = blue,
} 

\begin{document}

\title{Antiskyrmions and Bloch Skyrmions in Magnetic Dresselhaus Metals}

\author{Deepak S. Kathyat}
\author{Arnob Mukherjee}
\author{Sanjeev Kumar}

\address{ Department of Physical Sciences,
Indian Institute of Science Education and Research Mohali, Sector 81, S.A.S. Nagar, Manauli PO 140306, India \\
}

\begin{abstract}
We present a microscopic electronic description of how antiskyrmions can be stabilized in a Dresselhaus spin-orbit coupled magnetic metal. Furthermore, we show that the antiskyrmions can be tuned into Bloch skyrmions via a change in sign of hopping integral. The results are based on the state of the art hybrid Monte Carlo simulations. Origin of such topological textures is understood via an effective spin-only model. Our results uncover a novel connection between two very distinct topological spin textures of immense current interest, and present a microscopic explanation of skyrmion formation reported recently in certain magnetic Weyl semimetals.
\end{abstract}
\date{\today}

\maketitle

{\it Introduction:--}
The importance of magnetic skyrmion and antiskyrmion (ASk) quasiparticles in future technologies has been very well accepted \cite{Fert2017, Wiesendanger2016, Fert2013, Nagaosa2013b, Bogdanov2020}. Consequently, identifying new materials that host such exotic magnetic textures has become highly active and important area of research. Appearance of skyrmions has been confirmed with small angle neutron scattering and Lorentz transmission electron microscopy measurements in bulk as well as in thin films of a variety of chiral magnets \cite{Yu2011, Meyer2019, Hsu2018, Tonomura2012, Yu2018, Nagase2019,Romming2013b, Yu2012, Zhao2016}. On the other hand, antiskyrmions have been observed in bulk materials \cite{Nayak2017, Vir2019, jena2020observation} but not in thin films. While the early theoretical understanding of these textures was largely based on phenomenological magnetic models, recent years have witnessed efforts towards a microscopic electronic description of these topological textures \cite{Hayami2018,Hayami2019,Wang2020,Bezvershenko2018,hayami2021field,jin2021field}. Most of these recent studies focus on identifying the correct electronic model to stabilize one specific type of magnetic textures, such as, Neel- or Bloch-skyrmions or antiskyrmions.  

The role of Dzyaloshinskii-Moriya (DM) interactions in stabilizing skyrmion-like textures is also well documented \cite{chen2016,roessler,mohanta2019,iwasaki2014,Yi2009}. Current theories suggest that while skyrmions are stabilized by isotropic DM coupling, stabilization of antiskyrmions require anisotropic DM interactions \cite{hoffmann2017antiskyrmions,huang2017stabilization,qiu2020dynamics} or dipolar interactions \cite{koshibae2016theory}. At a more fundamental level, the DM interactions emerge from the spin-orbit coupling (SOC) which can be of atomic, Rashba or Dresselhaus type. Recent investigations have clarified how even in metallic systems the Rashba-type SOC generates terms that resemble the DM interactions. This has led to an elementary understanding of formation of Neel-type skyrmions in magnetic metals that support a moderate strength of Rashba SOC  \cite{Kathyat2020a,Kathyat2021}. However, a large number of magnetic metals with broken inversion symmetry are known to host Bloch-type skyrmions. In addition to a phenomenological understanding, microscopic theories based on geometrical frustration can also explain the formation of Bloch skyrmions \cite{PhysRevLett.108.017206}. However, not all materials that exhibit Bloch-skyrmion signatures possess a geometrically frustrated magnetic lattice. Furthermore, most of these materials are metals where a microscopic picture ignoring electronic itinerancy is highly questionable. 
Observations of skyrmions have also been reported in magnetic Weyl semimetals, which show nodal-point structure in the electronic bands as well as magnetic ordering \cite{araki2020magnetic,wu2020observation,PhysRevLett.124.017202,PhysRevB.102.184407}. Properties of the Weyl electrons in such materials can get enriched in the presence of skyrmion like magnetic textures in real space. The microscopic mechanism for the formation of these textures in Weyl semimetals is currently unknown. 

In this Letter, we show that Dresselhaus double exchange (DDE) model microscopically describes the formation of antiskyrmions. We study the model on
a square lattice via state-of-the-art hybrid Monte-Carlo (HMC) simulations. We find that the DDE model in the presence of Zeeman coupling to an external magnetic field stabilizes antiskyrmion textures. We then show that a simple change in sign of one of the hopping parameters converts antiskyrmions into Bloch skyrmions. This is physically relevant as many of the magnetic materials hosting Bloch skyrmions display the minimum location away from the $\Gamma$ point confirming a relative sign-difference between hopping parameters along different directions \cite{PhysRevB.93.195101,PhysRevB.99.134437,PhysRevLett.125.117204,koretsune2015control}. An effective spin-model is studied for a conventional understanding of these results. We further establish a connection between the DDE model and the magnetic Weyl semimetals, thereby explaining the recent experimental observations of skyrmions in the latter.

{\it Antiskyrmions in a model for Dresselhaus magnetic metals:--}
As a prototype of metallic magnets that lack inversion center in their structure, we consider a tight-binding model with Dresselhaus SOC and Hund's coupling. The Hamiltonian is given by,
\begin{eqnarray}
	H & = & - t \sum_{\langle ij \rangle,\sigma} (c^\dagger_{i\sigma} c^{}_{j\sigma} + {\textrm H.c.}) 
	+ \lambda \sum_{i} [\textrm{i} (c^{\dagger}_{i\downarrow} c^{}_{i+x\uparrow} + c^{\dagger}_{i\uparrow} c^{}_{i+x\downarrow}) \nonumber \\
	& & +  (c^{\dagger}_{i \downarrow} c^{}_{i+y\uparrow} - c^{\dagger}_{i\uparrow} c^{}_{i+y\downarrow})+ {\textrm H.c.}] - J_H \sum_{i} {\bf S}_i \cdot {\bf s}_i.
	\label{eq:Ham_DE}
\end{eqnarray}
Here, $c_{i\sigma} (c_{i\sigma}^\dagger$) annihilates (creates) an electron
at site ${i}$ with spin $\sigma$, $\langle ij \rangle$ implies that $i$ and $j$ are nearest neighbor (nn) sites. $\lambda$ denotes the strength of Dresselhaus SOC.
The last term describes a coupling between electronic spin ${\bf s}_i$ and localized core spins    ${\bf S}_i$. In many material involving Mn, Fe or rare earth ions a well localized core spin explicitly exists, and therefore the last term is justified. However, the scope of this term is broader since the Hubbard model at the mean-field level also generates an effective Hund's coupling \cite{PhysRevLett.101.156402,PhysRevB.93.195110}.
We parameterize $t = (1-\alpha) t_0$ and $\lambda = \alpha t_0$ and set $t_0=1$ as the reference energy scale.
Assuming large $J_H$ and taking the double-exchange approximation, we obtain the DDE Hamiltonian,
\begin{eqnarray}
	H_{\rm DDE} &=& \sum_{\langle ij \rangle, \gamma} [g^{\gamma}_{ij} d^{\dagger}_{i} d^{}_{j} + {\rm H.c.}] - h_z \sum_i S^z_i,
	\label{eq:Ham-DDE}
\end{eqnarray}
\noindent
where, $d^{}_{i} (d^{\dagger}_{i})$  annihilates (creates) an electron at site ${i}$ with spin parallel to the localized spin. 
The Zeeman coupling of spins to an external magnetic field of strength $h_z$ has also been included in Eq. (\ref{eq:Ham-DDE}).
Site $j = i + \gamma$ is the nn of site $i$ along spatial direction $\gamma = x,y$. The projected hopping $g^{\gamma}_{ij} = t^{\gamma}_{ij} + \lambda^{\gamma}_{ij}$ depend on the orientations of the local moments ${\bf S}_i$ and ${\bf S}_j$,
\begin{eqnarray}
	t^{\gamma}_{ij} & = & -t \big[\cos(\frac{\theta_i}{2}) \cos(\frac{\theta_j}{2}) 
	+ \sin(\frac{\theta_i}{2})  \sin(\frac{\theta_j}{2})e^{-\textrm{i} (\phi_i-\phi_j)} \big],
	\nonumber \\ 
	\lambda_{{ij}}^x & = & \textrm{i} \lambda \big[\sin(\frac {\theta_i}{2})  \cos(\frac {\theta_j}{2})e^{-\textrm{i} \phi_i} + \cos(\frac {\theta_i}{2})  \sin(\frac {\theta_j}{2})e^{\textrm{i} \phi_j}\big],
	\nonumber \\
	\lambda_{{ij}}^{y} & = & \lambda \big[\sin(\frac {\theta_i}{2})  \cos(\frac {\theta_j}{2})e^{-\textrm{i} \phi_i} - \cos(\frac {\theta_i}{2})  \sin(\frac {\theta_j}{2})e^{\textrm{i} \phi_j}\big],
	\label{eq:PH_DDE}
\end{eqnarray}
\noindent 
where, $\theta_i$ ($\phi_i$) denotes the polar (azimuthal) angle for localized spin ${\bf S}_i$.

We study the DDE Hamiltonian using numerically exact HMC simulations \cite{kumar2006,mukherjee2015,Kathyat2021}. Presence of skyrmions or antiskyrmions is inferred via local skyrmion density \cite{chen2016},
\begin{eqnarray} 
	\chi_{i} & = & \frac{1}{8\pi} [ {\bf S}_i \cdot ({\bf S}_{i+x} \times {\bf S}_{i+y} ) + {\bf S}_i \cdot ({\bf S}_{i-x} \times {\bf S}_{i-y})].
\end{eqnarray}
\noindent
Total skyrmion density is defined as, $\chi = \sum_i \chi_i$.  
We also compute the relevant component of vector chirality $\eta$ as, 
\begin{eqnarray} 
	\eta & = & \frac{1}{N} \sum_{i} ({\bf S}_{i} \times {\bf S}_{i+y} )\cdot \hat{y} -({\bf S}_{i} \times {\bf S}_{i+x} )\cdot \hat{x}.
	\label{eq:vc_DDE}
\end{eqnarray} 
\begin{figure}[t]
	\includegraphics[width=1.0 \columnwidth,angle=0,clip=true]{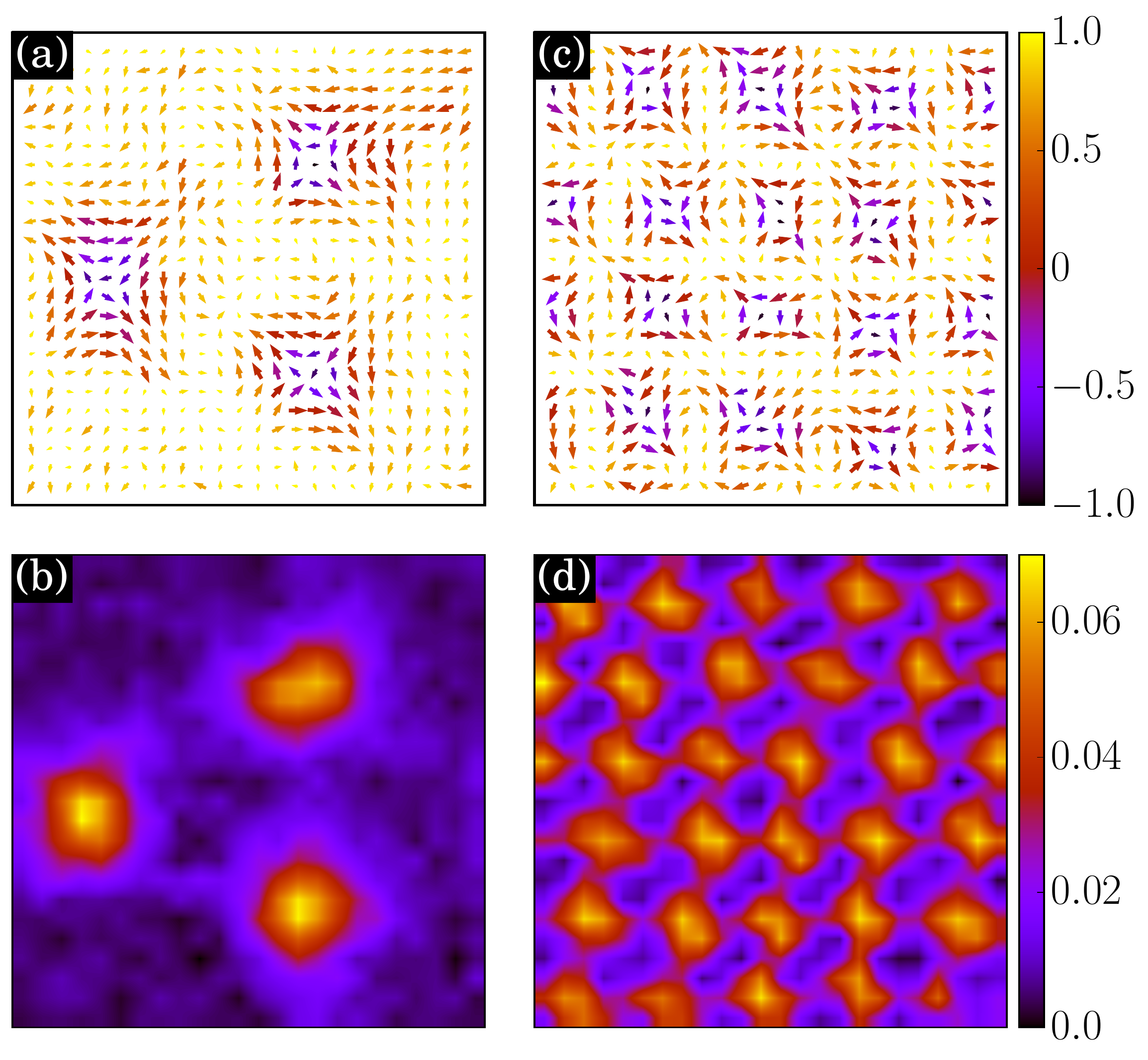} 
	\caption{Snapshots of spin configurations, (a), (c), and the local skyrmion density, (b), (d), at $T=0.01$ for representative values of $\alpha$ and $h_z$ as obtained in HMC simulations: (a)-(b) $\alpha =0.25 $, $h_z=0.03$; (c)-(d) $\alpha=0.4$, $h_z=0.06$. The HMC simulations are  performed  on  $24 \times 24$ lattice for an average band filling fraction of $0.3$.
	}
	\label{tca}
\end{figure}

Results obtained via HMC simulations for two representative values of $\alpha$, at electronic filling of $n = 0.3$, are shown in Fig. \ref{tca}. The existence of antiskyrmions in the DDE Hamiltonian is explicitly demonstrated via the spin configurations (see Fig. \ref{tca}(a), (c)) as well as skyrmion density maps  (see Fig. \ref{tca}(b), (d)) at low temperatures representative of the ground states. The positive sign of $\chi$ that is opposite to sign of the polarity (defined by magnetization of the central spin of the texture) reveals that the ground states are antiskyrmions. We find that small values of $\alpha$ lead to sparse antiskyrmions within the zero field cooled (ZFC) protocol, and the packing (size) of antiskyrmions increases (decreases) with increasing $\alpha$.
Note that the Hamiltonian Eq. (\ref{eq:Ham-DDE}) does not contain any direct spin-spin interaction terms. Therefore, the magnetic patterns discussed above are stabilized by interactions that are mediated via conduction electrons. A conventional understanding of the formation of such textures can only emerge via identification of the emergent spin-spin interactions in this model. To this end, we now present results on an effective spin-only model derived from the DDE Hamiltonian.

{\it Antiskyrmions in the effective spin model:--}
We derive an effective spin model for $H_{\rm DDE}$ in similar way as derived for $H_{\rm RDE}$ in \cite{Kathyat2021}, we obtain,
\begin{eqnarray}   \label{eq:ESH_DDE} 
	H_{\rm eff} & = &-\sum_{\langle ij \rangle, \gamma} D^{x(y)}_{ij} f^{x(y)}_{ij} - h_z \sum_i S^z_i,  \nonumber \\
	\sqrt{2} f^{x(y)}_{ij} & = &  \Big[ t^2\{1+{\bf S}_i \cdot {\bf S}_j\} -(+) 2t\lambda \hat{x} (\hat{y}) \cdot \{{\bf S}_i \times{\bf S}_j\}  \nonumber  \\
	& & + \lambda^2\big \{1-{\bf S}_i \cdot {\bf S}_j+ 2\{\hat{x} (\hat{y}) \cdot {\bf S}_i\}\{\hat{x} (\hat{y}) \cdot {\bf S}_j\} \big\} \Big]^{1/2}, \nonumber \\
	D^{x(y)}_{ij} & = & \langle [e^{{\rm i} h^{x(y)}_{ij}} d^{\dagger}_{i} d^{}_{j} + {\textrm H.c.}] \rangle_{gs}.
\end{eqnarray}

\noindent
In the above, $f^{x(y)}_{ij}$ ($h^{x(y)}_{ij}$) is the modulus (argument) of complex number $g^{x(y)}_{ij}$ along $x (y)$ direction. $\langle \hat{O} \rangle_{gs}$ denotes expectation values of operator $\hat{O}$ in the ground state.
It has been shown that using a constant value of $D^{x(y)}_{ij}$ captures the essential physics of the Hamiltonian Eq. (\ref{eq:ESH_DDE}), therefore we set $D^{x(y)}_{ij} \equiv D_0 = 1$ and simulate $H_{\rm eff}$ using the conventional classical MC \cite{Kathyat2020a,Kathyat2021}. The first term under the square-root in the definition of $\sqrt{2} f^{x(y)}_{ij}$ is an isotropic term, favoring ferromagnetic ordering, the second term is an anisotropic DM-like exchange coupling of strength $-2t\lambda$ ($+2t\lambda$) along $x$ ($y$) direction and the last term is a pseudo-dipolar term that leads to magnetic groundstate degeneracy of non-geometrical origin.
 
\begin{figure}[t]
	\includegraphics[width=1.0 \columnwidth,angle=0,clip=true]{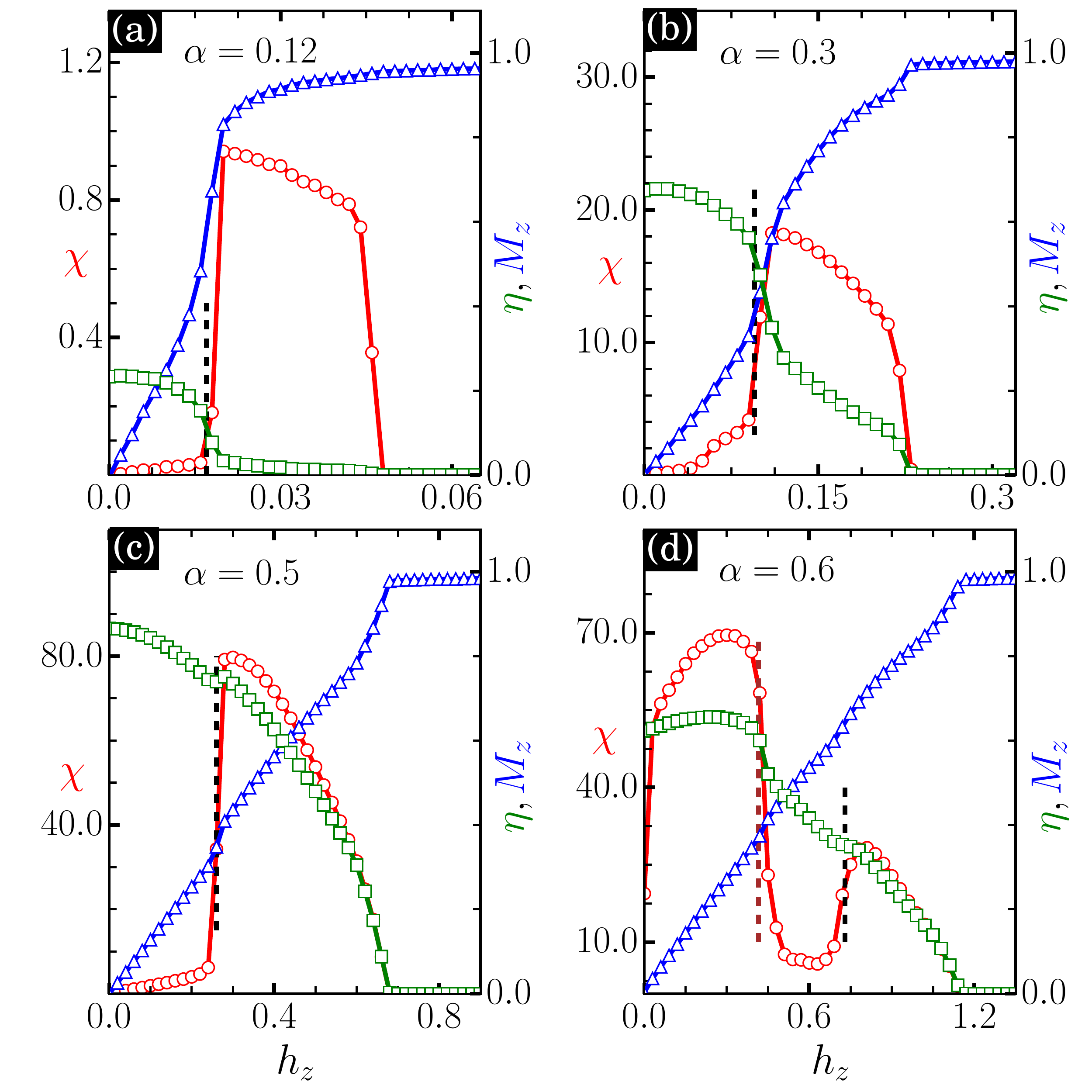}  
	\caption{(a) - (d) Variation of magnetization $M_z$ (triangles), total skyrmion density $\chi$ (circles) and vector chirality $\eta$ (squares) with increasing $h_z$ for different values of $\alpha$.
	}
	\label{order_para}
\end{figure} 

In Fig. \ref{order_para} we show the field-dependence of magnetization, $M_z = \frac{1}{N} \sum_i S^z_i$, vector chirality $\eta$ and skyrmion density $\chi$ obtained via simulations of $H_{\rm eff}$ Eq. (\ref{eq:ESH_DDE}). 
For small values of $\alpha$, a linear increase of magnetization for small $h_z$, is followed by a non-linear behavior that is accompanied by a sharp increase in $\chi$ (see Fig. \ref{order_para}(a), (b)). We can infer that the emergence of antiskyrmions makes it difficult for spins to align along the direction of the external magnetic field. 
Existence of a finite $\eta$ in the absence of magnetic field can be understood from the DM-like terms in $H_{\rm eff}$.
Upon increasing $h_z$, $\eta$ displays a sharp decrease concomitant with the increase in $\chi$ (see Fig. \ref{order_para}(a), (b)). Finally, as the system approaches a field-enforced saturated FM state, both $\chi$ and $\eta$ vanish. For $\alpha = 0.5$, the change in $\chi$ near $h_z = 0.25$ is accompanied by a weak discontinuity in both $M_z$ and $\eta$ (see Fig. \ref{order_para}(c)). This qualitatively different behavior is an indicator of packed ASk state, as will be illustrated below with the help of real space spin configurations.
For $\alpha = 0.6$, $\chi$ is finite even at $h_z = 0$ indicating the existence of a zero-field antiskyrmion crystal state. A sharp reduction in $\chi$ near $h_z = 0.4$ indicates the destabilization of this low-field ASk state. At higher fields a qualitatively different finite-field ASk state appears again before the system approaches a saturated FM state (see Fig. \ref{order_para}(d)).

\begin{figure}[t]
	\includegraphics[width=1.0 \columnwidth,angle=0,clip=true]{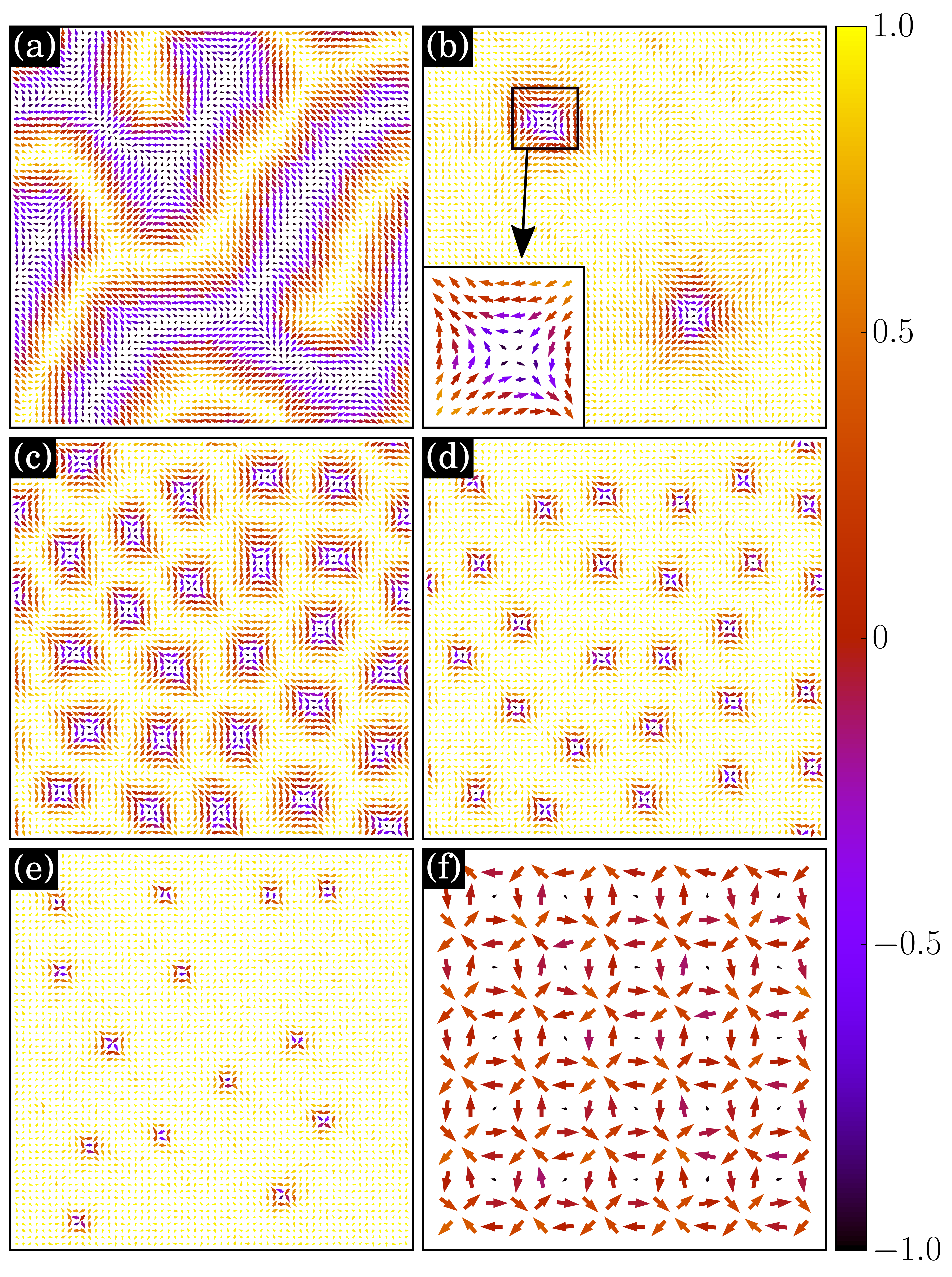}   
	\caption{Snapshots of spin configurations at low temperature for representative values of $\alpha$ and $h_z$. (a) filamentary domain walls at $\alpha=0.15$, $h_z=0$, (b) sparse antiskyrmions at $\alpha=0.15$, $h_z=0.036$, (c) packed antiskyrmions at $\alpha=0.3$, $h_z=0.11$, (d) ASk size reduction at $\alpha=0.3$, $h_z=0.18$ w.r.t. (c), (e) ASk number reduction  at $\alpha=0.3$, $h_z=0.22$ w.r.t. (d), and (f) the zero-field ASk crystal at $\alpha=0.6$. For (a)-(e) $60 \times 60$ lattices are shown. For (f) a smaller section, $16 \times 16$ of the full lattice is shown for clarity. The color bar corresponds to the $z$ component of spins.
	}
	\label{spin_confi}
\end{figure} 

We show in Fig. \ref{spin_confi} the evolution of magnetic textures with change in $\alpha$ and $h_z$ within  $H_{\rm eff}$. Within the ZFC protocol at finite temperatures, we find states with filamentary domain wall structure characterized by the diffuse ring pattern in the spin structure factor  \cite{Kathyat2020a,Kathyat2021} in the absence of external field for small $\alpha$ (see Fig. \ref{spin_confi}(a)). The filamentary domain walls are randomly oriented and look similar to those reported in the Rashba model \cite{Kathyat2020a}. However, unlike Rashba systems wherein a spiral in xz (yz) plane is preferred along x (y) direction, in Dresselhaus metals a spiral in xz (yz) plane along y (x) direction is favored. These states can be viewed as the parent state for the antiskyrmions when magnetic field is applied (see Fig. \ref{spin_confi}(b)). For larger values of $\alpha$, SQ spiral state gives way to the packed ASk phase (see Fig. \ref{spin_confi}(c)). For a given $\alpha$, increasing $h_z$ leads to a reduction of the size by polarizing the spins at the edges of skyrmions (compare Fig. \ref{spin_confi} (c) and (d)), followed by a reduction in the number of antiskyrmions (compare Fig. \ref{spin_confi} (d) and (e)). A perfectly ordered crystal of smallest possible antiskyrmions on a square lattice is obtained in the absence of external field at $\alpha = 0.6$ (see Fig. \ref{spin_confi} (f)). 

{\it Turning antiskyrmions into Bloch skyrmions:--}
Our results so far show that antiskyrmions exist in Dresselhaus metals, and Neel-type skyrmions in Rashba metals \cite{Kathyat2021}. However, Bloch-type skyrmions have been reported in many  
magnetic metals. Our description will certainly remain incomplete if we cannot show the formation of Bloch-type skyrmions within our microscopic model. We now discuss an elegant and physically relevant mechanism to turn antiskyrmions into Bloch skyrmions. 

\begin{figure}[h]
	\includegraphics[width=1.0 \columnwidth,angle=0,clip=true]{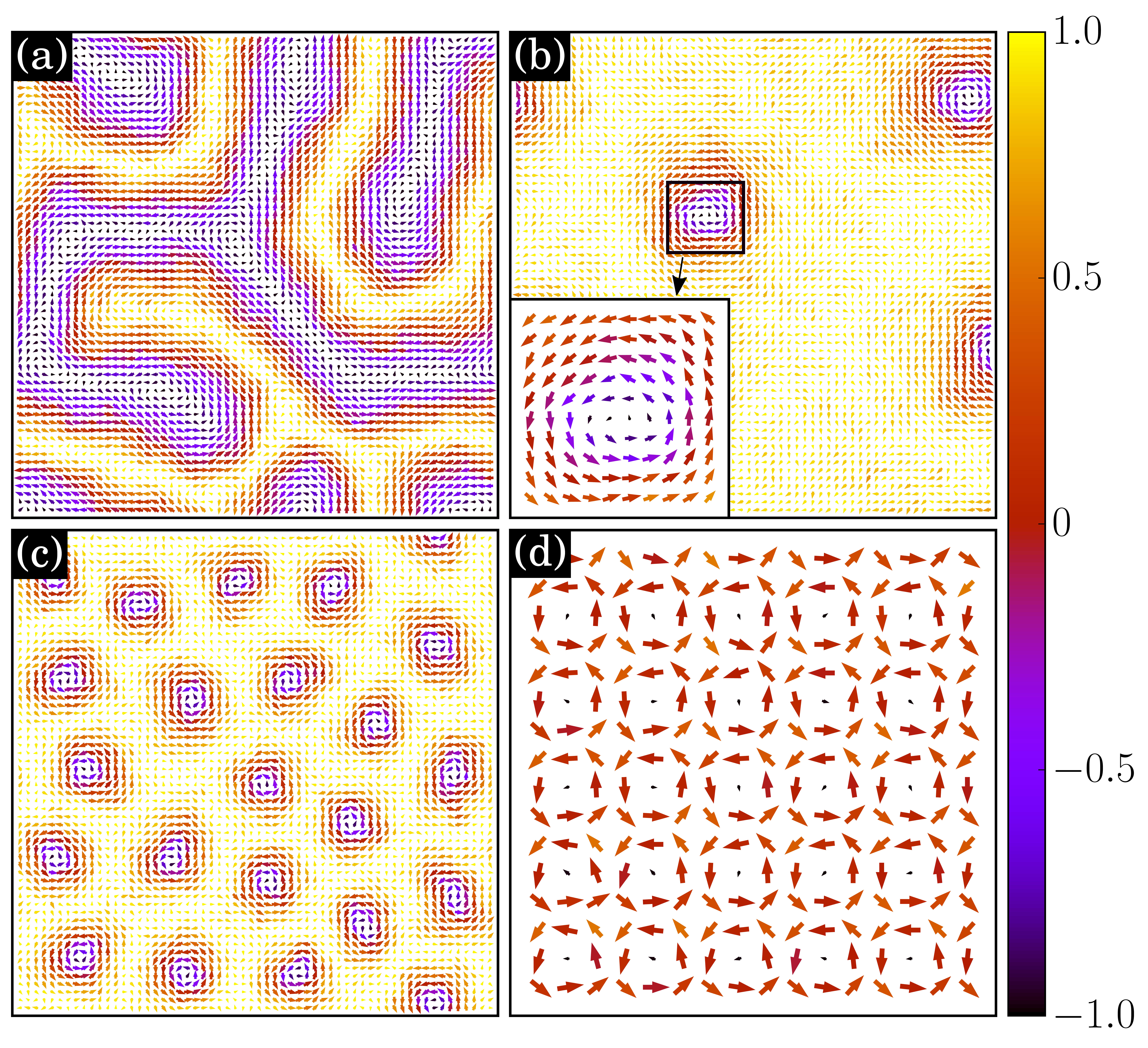}  
	\caption{Low temperature snapshots of ground states in modified model for representative values of $\alpha$ and $h_z$. (a) filamentary domain wall state at $\alpha=0.16$, $h_z=0$, (b) sparse Bloch skyrmions at $\alpha=0.16$, $h_z=0.036$, (c) packed Bloch skyrmions at $\alpha=0.3$, $h_z=0.12$ and (d) Bloch skyrmions crystal at $\alpha=0.6$, $h_z=0.0$.
	}
	\label{bloch}
\end{figure}

Introduction of a relative sign between x and y direction hopping parameter modifies the 
$f^{\gamma}_{ij}$ of the effective Hamiltonian Eq. (\ref{eq:ESH_DDE}) as follows,
\begin{eqnarray}   \label{eq:MESH}    
	\sqrt{2} f^{\gamma}_{ij} & = &  \big[ t^2(1+{\bf S}_i \cdot {\bf S}_j) + 2t\lambda  \hat{\gamma} \cdot ({\bf S}_i \times{\bf S}_j)  \nonumber  \\
	& & + \lambda^2(1-{\bf S}_i \cdot {\bf S}_j+2 (\hat{\gamma} \cdot {\bf S}_i)(\hat{\gamma} \cdot {\bf S}_j)) \big]^{1/2}, \nonumber \\
\end{eqnarray} 
\noindent
where, $\hat{\gamma} \in \{x, y\}$. We simulate the modified $H_{\rm eff}$  Eq. (\ref{eq:MESH}) using the conventional classical MC. In Fig. \ref{bloch}  we show the evolution of magnetic textures with change in $\alpha$ and $h_z$ within modified $H_{\rm eff}$. We find that, the domain junctions in the filamentary domain wall states for small $\alpha$ (see Fig. \ref{bloch}(a)) become nucleation centers for  Bloch skyrmions when magnetic field is applied (see Fig. \ref{bloch}(b)). For larger values of $\alpha$, packed Bloch skyrmion phase (see Fig. \ref{bloch}(c)) is stabilized as the ground state. While the packed Bloch skyrmions exist in the presence of external field, at $\alpha = 0.6$ the square-lattice Bloch SkX is the ground state already at $h_z = 0$ (see Fig. \ref{bloch} (d)). In summary, the evolution of the phases remains the same as that discussed in previous section for antiskyrmions and the nature of textures gets converted into Bloch-type skyrmions.

Interestingly, there is a simple way to find out if the effective tight-binding model for a given metal has a relative sign difference between hopping parameters along different directions. If the maximum or the minimum of the relevant energy bands that cross the Fermi level lies at the $\Gamma$ point, then all hoppings have same sign. This observation, together with the results discussed so far, leads to a very important general conjecture. 
Materials that host Bloch skyrmions should not have the maximum or minimum of the relevant bands without SOC at the $\Gamma$ point . We found many examples where this is indeed true  \cite{PhysRevB.93.195101,PhysRevB.99.134437,PhysRevLett.125.117204,koretsune2015control}. Similarly, Neel skyrmions are more likely when the maximum or the minimum lies at the $\Gamma$ point, and on surfaces or interfaces so that Rashba effect is dominant. Once again, this condition seems to hold for many Neel skyrmion hosts \cite{PhysRevB.92.121112,bhowal2019electric,PhysRevB.99.245145}.  

\begin{figure}[h]
	\includegraphics[width=1.0 \columnwidth,angle=0,clip=true]{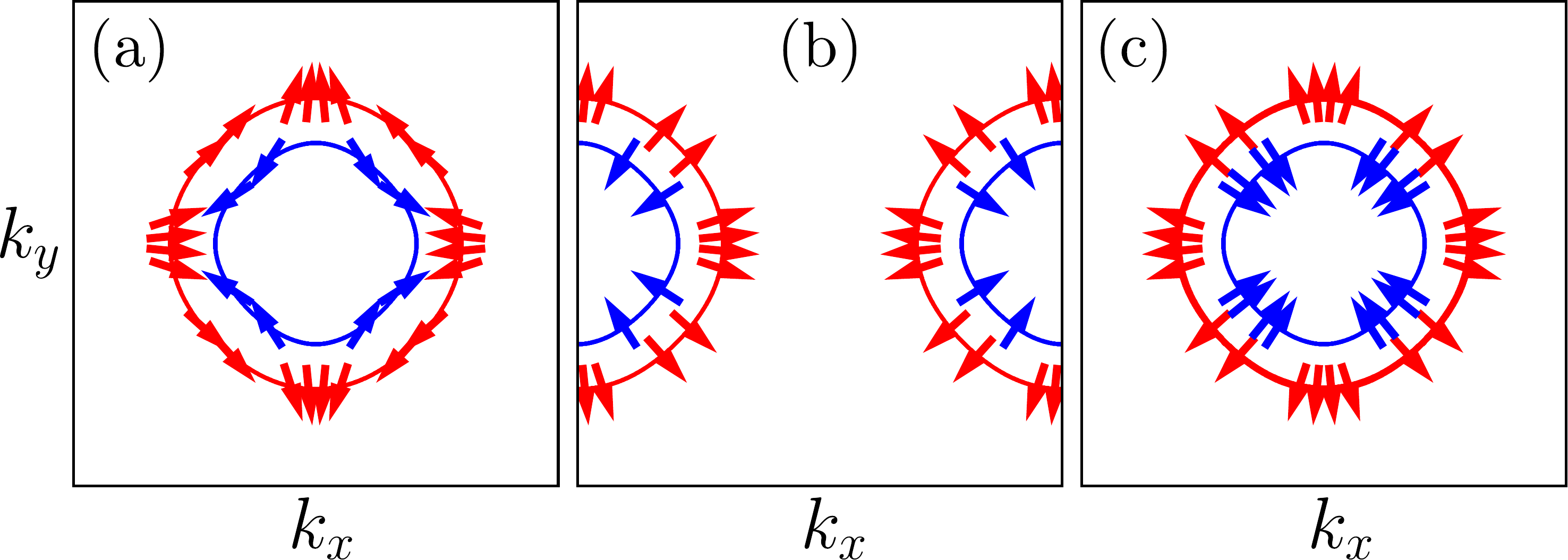}
	\caption{Spin textures in the momentum space on Fermi contours at $1/5$ filling for (a) Dresselhaus SOC strength $\lambda = 0.3$ and $t_{x} = t_{y} = 1.0$ , (b) $\lambda = 0.3$, $t_{x}=-1.0, t_{y}=1.0$, and (c) Dresselhaus SOC strength $\lambda$ having an extra minus sign along the x direction leading to a lattice version of the $\boldsymbol{\sigma \cdot k}$ model with $t_{x} = t_{y} = 1.0$.
	In all panels only the first Brillouin zones are displayed.}
	\label{fermi_contour}
\end{figure}

Finally, we discuss a connection with skyrmions in Weyl semimetals. The Hamiltonian Eq. 
(\ref{eq:Ham_DE}) for $J_{\rm H} = 0$ can be easily diagonalized. The spin-momentum locking 
leads to well defined spin-textures in the momentum space (see Fig. \ref{fermi_contour}).
From our results we conclude that the nature of momentum-space spin textures completely determines the nature of real space magnetic texture that appear when Hund's or Hubbard interactions are included. Spin textures on the Fermi contours for Hamiltonian Eq. \ref{eq:Ham_DE} at $J_{\rm H} = 0$ display a specific antiskyrmion-like pattern (see Fig. \ref{fermi_contour} (a)). If the sign of one of the hopping terms is reversed then the momentum-space spin textures attain a Neel-skyrmion-like pattern, but away from the $\Gamma$ point (see Fig. \ref{fermi_contour} (b)). Finally, introducing a relative sign in the SOC coupling $\lambda$ along $x$ and $y$ directions leads to a Neel-skyrmion-like pattern centered at the $\Gamma$ point (see Fig. \ref{fermi_contour} (c)).
In all cases, the real-space magnetic textures are such that the spin directions around the core are orthogonal to the momentum space spin directions. 
The effective Hamiltonian Eq. (\ref{eq:ESH_DDE}) and Eq. (\ref{eq:MESH}) provide a simple explanation for this fact. Since the DM-like term is proportional to $t \lambda$, change in sign of either of these parameters will lead to the identical result for the magnetic patterns.
We note that a change in sign of SOC strength $\lambda$ along one direction transform the continuum version of the Dresselhaus Hamiltonian $\lambda(\sigma^x k_x - \sigma^y k_y)$-form to Weyl semimetal $\lambda(\boldsymbol{\sigma \cdot k})$. Therefore, the Weyl semimetals that are described by $\boldsymbol{\sigma \cdot k}$ Hamiltonian should naturally lead to Bloch skyrmions \cite{araki2020magnetic}. Therefore, our results are directly relevant for understanding the magnetism of many magnetic Weyl semimetals.

{\it Conclusion:--}
A number of antiskyrmion host materials, such as $\text{Mn}_{1.4}\text{Pt}_{0.9}\text{Pd}_{0.1}\text{Sn}$, $\text{Mn}_{1.4}\text{Pt}\text{Sn}$ and $\text{Mn}_{2}\text{Rh}_{0.95}\text{Ir}_{0.05}\text{Sn}$, happen to be inverse Heusler metals lacking a center of inversion \cite{Nayak2017, Vir2019, jena2020observation}. Presence of a magnetic ion together with heavy elements suggests that an appropriate description of magnetism of these materials can be in terms of a double exchange model modified by Dresselhaus SOC. We establish this by explicitly showing that magnetic Dresselhaus metals support nanoscale antiskyrmions. 
An understanding of stability of antiskyrmions is achieved in terms of a conventional spin-only model derived from the microscopic DDE Hamiltonian. We provide a recipe for turning antiskyrmions into Bloch skyrmions, and thereby providing a novel connection between the type of topological magnetic textures and the underlying electronic band structure. 
Such a connection will be completely missed in a spin-only model written without reference to a starting microscopic model. We also discuss the connection between the $\boldsymbol{\sigma \cdot k}$ model typically used to describe Weyl semimetals and the hopping-sign modified Dresselhaus Hamiltonian. This connection immediately provides an explanation for the existence of Bloch skyrmions in magnetic Weyl semimetals.

{\it Acknowledgments:--}
We acknowledge the use of computing facility at IISER Mohali.

\bibliographystyle{apsrev4-1} 

\begin{thebibliography}{50}%
\makeatletter
\providecommand \@ifxundefined [1]{%
 \@ifx{#1\undefined}
}%
\providecommand \@ifnum [1]{%
 \ifnum #1\expandafter \@firstoftwo
 \else \expandafter \@secondoftwo
 \fi
}%
\providecommand \@ifx [1]{%
 \ifx #1\expandafter \@firstoftwo
 \else \expandafter \@secondoftwo
 \fi
}%
\providecommand \natexlab [1]{#1}%
\providecommand \enquote  [1]{``#1''}%
\providecommand \bibnamefont  [1]{#1}%
\providecommand \bibfnamefont [1]{#1}%
\providecommand \citenamefont [1]{#1}%
\providecommand \href@noop [0]{\@secondoftwo}%
\providecommand \href [0]{\begingroup \@sanitize@url \@href}%
\providecommand \@href[1]{\@@startlink{#1}\@@href}%
\providecommand \@@href[1]{\endgroup#1\@@endlink}%
\providecommand \@sanitize@url [0]{\catcode `\\12\catcode `\$12\catcode
  `\&12\catcode `\#12\catcode `\^12\catcode `\_12\catcode `\%12\relax}%
\providecommand \@@startlink[1]{}%
\providecommand \@@endlink[0]{}%
\providecommand \url  [0]{\begingroup\@sanitize@url \@url }%
\providecommand \@url [1]{\endgroup\@href {#1}{\urlprefix }}%
\providecommand \urlprefix  [0]{URL }%
\providecommand \Eprint [0]{\href }%
\providecommand \doibase [0]{http://dx.doi.org/}%
\providecommand \selectlanguage [0]{\@gobble}%
\providecommand \bibinfo  [0]{\@secondoftwo}%
\providecommand \bibfield  [0]{\@secondoftwo}%
\providecommand \translation [1]{[#1]}%
\providecommand \BibitemOpen [0]{}%
\providecommand \bibitemStop [0]{}%
\providecommand \bibitemNoStop [0]{.\EOS\space}%
\providecommand \EOS [0]{\spacefactor3000\relax}%
\providecommand \BibitemShut  [1]{\csname bibitem#1\endcsname}%
\let\auto@bib@innerbib\@empty
\bibitem [{\citenamefont {Fert}\ \emph {et~al.}(2017)\citenamefont {Fert},
  \citenamefont {Reyren},\ and\ \citenamefont {Cros}}]{Fert2017}%
  \BibitemOpen
  \bibfield  {author} {\bibinfo {author} {\bibfnamefont {A.}~\bibnamefont
  {Fert}}, \bibinfo {author} {\bibfnamefont {N.}~\bibnamefont {Reyren}}, \ and\
  \bibinfo {author} {\bibfnamefont {V.}~\bibnamefont {Cros}},\ }\href {\doibase
  10.1038/natrevmats.2017.31} {\bibfield  {journal} {\bibinfo  {journal} {Nat.
  Rev. Mater.}\ }\textbf {\bibinfo {volume} {2}},\ \bibinfo {pages} {17031}
  (\bibinfo {year} {2017})}\BibitemShut {NoStop}%
\bibitem [{\citenamefont {Wiesendanger}(2016)}]{Wiesendanger2016}%
  \BibitemOpen
  \bibfield  {author} {\bibinfo {author} {\bibfnamefont {R.}~\bibnamefont
  {Wiesendanger}},\ }\href {https://doi.org/10.1038/natrevmats.2016.44}
  {\bibfield  {journal} {\bibinfo  {journal} {Nature Reviews Materials}\
  }\textbf {\bibinfo {volume} {1}},\ \bibinfo {pages} {1} (\bibinfo {year}
  {2016})}\BibitemShut {NoStop}%
\bibitem [{\citenamefont {Fert}\ \emph {et~al.}(2013)\citenamefont {Fert},
  \citenamefont {Cros},\ and\ \citenamefont {Sampaio}}]{Fert2013}%
  \BibitemOpen
  \bibfield  {author} {\bibinfo {author} {\bibfnamefont {A.}~\bibnamefont
  {Fert}}, \bibinfo {author} {\bibfnamefont {V.}~\bibnamefont {Cros}}, \ and\
  \bibinfo {author} {\bibfnamefont {J.}~\bibnamefont {Sampaio}},\ }\href
  {\doibase 10.1038/nnano.2013.29} {\bibfield  {journal} {\bibinfo  {journal}
  {Nat. Nanotechnol.}\ }\textbf {\bibinfo {volume} {8}},\ \bibinfo {pages}
  {152} (\bibinfo {year} {2013})}\BibitemShut {NoStop}%
\bibitem [{\citenamefont {Nagaosa}\ and\ \citenamefont
  {Tokura}(2013)}]{Nagaosa2013b}%
  \BibitemOpen
  \bibfield  {author} {\bibinfo {author} {\bibfnamefont {N.}~\bibnamefont
  {Nagaosa}}\ and\ \bibinfo {author} {\bibfnamefont {Y.}~\bibnamefont
  {Tokura}},\ }\href {\doibase 10.1038/nnano.2013.243} {\bibfield  {journal}
  {\bibinfo  {journal} {Nat. Nanotechnol.}\ }\textbf {\bibinfo {volume} {8}},\
  \bibinfo {pages} {899} (\bibinfo {year} {2013})}\BibitemShut {NoStop}%
\bibitem [{\citenamefont {Bogdanov}\ and\ \citenamefont
  {Panagopoulos}(2020)}]{Bogdanov2020}%
  \BibitemOpen
  \bibfield  {author} {\bibinfo {author} {\bibfnamefont {A.~N.}\ \bibnamefont
  {Bogdanov}}\ and\ \bibinfo {author} {\bibfnamefont {C.}~\bibnamefont
  {Panagopoulos}},\ }\href {\doibase 10.1063/PT.3.4431} {\bibfield  {journal}
  {\bibinfo  {journal} {Phys. Today}\ }\textbf {\bibinfo {volume} {73}},\
  \bibinfo {pages} {44} (\bibinfo {year} {2020})}\BibitemShut {NoStop}%
\bibitem [{\citenamefont {Yu}\ \emph {et~al.}(2011)\citenamefont {Yu},
  \citenamefont {Kanazawa}, \citenamefont {Onose}, \citenamefont {Kimoto},
  \citenamefont {Zhang}, \citenamefont {Ishiwata}, \citenamefont {Matsui},\
  and\ \citenamefont {Tokura}}]{Yu2011}%
  \BibitemOpen
  \bibfield  {author} {\bibinfo {author} {\bibfnamefont {X.~Z.}\ \bibnamefont
  {Yu}}, \bibinfo {author} {\bibfnamefont {N.}~\bibnamefont {Kanazawa}},
  \bibinfo {author} {\bibfnamefont {Y.}~\bibnamefont {Onose}}, \bibinfo
  {author} {\bibfnamefont {K.}~\bibnamefont {Kimoto}}, \bibinfo {author}
  {\bibfnamefont {W.~Z.}\ \bibnamefont {Zhang}}, \bibinfo {author}
  {\bibfnamefont {S.}~\bibnamefont {Ishiwata}}, \bibinfo {author}
  {\bibfnamefont {Y.}~\bibnamefont {Matsui}}, \ and\ \bibinfo {author}
  {\bibfnamefont {Y.}~\bibnamefont {Tokura}},\ }\href {\doibase
  10.1038/nmat2916} {\bibfield  {journal} {\bibinfo  {journal} {Nat. Mater.}\
  }\textbf {\bibinfo {volume} {10}},\ \bibinfo {pages} {106} (\bibinfo {year}
  {2011})}\BibitemShut {NoStop}%
\bibitem [{\citenamefont {Meyer}\ \emph {et~al.}(2019)\citenamefont {Meyer},
  \citenamefont {Perini}, \citenamefont {von Malottki}, \citenamefont
  {Kubetzka}, \citenamefont {Wiesendanger}, \citenamefont {von Bergmann},\ and\
  \citenamefont {Heinze}}]{Meyer2019}%
  \BibitemOpen
  \bibfield  {author} {\bibinfo {author} {\bibfnamefont {S.}~\bibnamefont
  {Meyer}}, \bibinfo {author} {\bibfnamefont {M.}~\bibnamefont {Perini}},
  \bibinfo {author} {\bibfnamefont {S.}~\bibnamefont {von Malottki}}, \bibinfo
  {author} {\bibfnamefont {A.}~\bibnamefont {Kubetzka}}, \bibinfo {author}
  {\bibfnamefont {R.}~\bibnamefont {Wiesendanger}}, \bibinfo {author}
  {\bibfnamefont {K.}~\bibnamefont {von Bergmann}}, \ and\ \bibinfo {author}
  {\bibfnamefont {S.}~\bibnamefont {Heinze}},\ }\href
  {https://doi.org/10.1038/s41467-019-11831-4} {\bibfield  {journal} {\bibinfo
  {journal} {Nature communications}\ }\textbf {\bibinfo {volume} {10}},\
  \bibinfo {pages} {1} (\bibinfo {year} {2019})}\BibitemShut {NoStop}%
\bibitem [{\citenamefont {Hsu}\ \emph {et~al.}(2018)\citenamefont {Hsu},
  \citenamefont {R{\'o}zsa}, \citenamefont {Finco}, \citenamefont {Schmidt},
  \citenamefont {Palot{\'a}s}, \citenamefont {Vedmedenko}, \citenamefont
  {Udvardi}, \citenamefont {Szunyogh}, \citenamefont {Kubetzka}, \citenamefont
  {von Bergmann} \emph {et~al.}}]{Hsu2018}%
  \BibitemOpen
  \bibfield  {author} {\bibinfo {author} {\bibfnamefont {P.-J.}\ \bibnamefont
  {Hsu}}, \bibinfo {author} {\bibfnamefont {L.}~\bibnamefont {R{\'o}zsa}},
  \bibinfo {author} {\bibfnamefont {A.}~\bibnamefont {Finco}}, \bibinfo
  {author} {\bibfnamefont {L.}~\bibnamefont {Schmidt}}, \bibinfo {author}
  {\bibfnamefont {K.}~\bibnamefont {Palot{\'a}s}}, \bibinfo {author}
  {\bibfnamefont {E.}~\bibnamefont {Vedmedenko}}, \bibinfo {author}
  {\bibfnamefont {L.}~\bibnamefont {Udvardi}}, \bibinfo {author} {\bibfnamefont
  {L.}~\bibnamefont {Szunyogh}}, \bibinfo {author} {\bibfnamefont
  {A.}~\bibnamefont {Kubetzka}}, \bibinfo {author} {\bibfnamefont
  {K.}~\bibnamefont {von Bergmann}},  \emph {et~al.},\ }\href
  {https://doi.org/10.1038/s41467-018-04015-z} {\bibfield  {journal} {\bibinfo
  {journal} {Nature communications}\ }\textbf {\bibinfo {volume} {9}},\
  \bibinfo {pages} {1} (\bibinfo {year} {2018})}\BibitemShut {NoStop}%
\bibitem [{\citenamefont {Tonomura}\ \emph {et~al.}(2012)\citenamefont
  {Tonomura}, \citenamefont {Yu}, \citenamefont {Yanagisawa}, \citenamefont
  {Matsuda}, \citenamefont {Onose}, \citenamefont {Kanazawa}, \citenamefont
  {Park},\ and\ \citenamefont {Tokura}}]{Tonomura2012}%
  \BibitemOpen
  \bibfield  {author} {\bibinfo {author} {\bibfnamefont {A.}~\bibnamefont
  {Tonomura}}, \bibinfo {author} {\bibfnamefont {X.}~\bibnamefont {Yu}},
  \bibinfo {author} {\bibfnamefont {K.}~\bibnamefont {Yanagisawa}}, \bibinfo
  {author} {\bibfnamefont {T.}~\bibnamefont {Matsuda}}, \bibinfo {author}
  {\bibfnamefont {Y.}~\bibnamefont {Onose}}, \bibinfo {author} {\bibfnamefont
  {N.}~\bibnamefont {Kanazawa}}, \bibinfo {author} {\bibfnamefont {H.~S.}\
  \bibnamefont {Park}}, \ and\ \bibinfo {author} {\bibfnamefont
  {Y.}~\bibnamefont {Tokura}},\ }\href {\doibase 10.1021/nl300073m} {\bibfield
  {journal} {\bibinfo  {journal} {Nano Lett.}\ }\textbf {\bibinfo {volume}
  {12}},\ \bibinfo {pages} {1673} (\bibinfo {year} {2012})}\BibitemShut
  {NoStop}%
\bibitem [{\citenamefont {Yu}\ \emph {et~al.}(2018)\citenamefont {Yu},
  \citenamefont {Koshibae}, \citenamefont {Tokunaga}, \citenamefont {Shibata},
  \citenamefont {Taguchi}, \citenamefont {Nagaosa},\ and\ \citenamefont
  {Tokura}}]{Yu2018}%
  \BibitemOpen
  \bibfield  {author} {\bibinfo {author} {\bibfnamefont {X.}~\bibnamefont
  {Yu}}, \bibinfo {author} {\bibfnamefont {W.}~\bibnamefont {Koshibae}},
  \bibinfo {author} {\bibfnamefont {Y.}~\bibnamefont {Tokunaga}}, \bibinfo
  {author} {\bibfnamefont {K.}~\bibnamefont {Shibata}}, \bibinfo {author}
  {\bibfnamefont {Y.}~\bibnamefont {Taguchi}}, \bibinfo {author} {\bibfnamefont
  {N.}~\bibnamefont {Nagaosa}}, \ and\ \bibinfo {author} {\bibfnamefont
  {Y.}~\bibnamefont {Tokura}},\ }\href
  {https://doi.org/10.1038/s41586-018-0745-3} {\bibfield  {journal} {\bibinfo
  {journal} {Nature}\ }\textbf {\bibinfo {volume} {564}},\ \bibinfo {pages}
  {95} (\bibinfo {year} {2018})}\BibitemShut {NoStop}%
\bibitem [{\citenamefont {Nagase}\ \emph {et~al.}(2019)\citenamefont {Nagase},
  \citenamefont {Komatsu}, \citenamefont {So}, \citenamefont {Ishida},
  \citenamefont {Yoshida}, \citenamefont {Kawaguchi}, \citenamefont {Tanaka},
  \citenamefont {Saitoh}, \citenamefont {Ikarashi}, \citenamefont {Kuwahara},\
  and\ \citenamefont {Nagao}}]{Nagase2019}%
  \BibitemOpen
  \bibfield  {author} {\bibinfo {author} {\bibfnamefont {T.}~\bibnamefont
  {Nagase}}, \bibinfo {author} {\bibfnamefont {M.}~\bibnamefont {Komatsu}},
  \bibinfo {author} {\bibfnamefont {Y.~G.}\ \bibnamefont {So}}, \bibinfo
  {author} {\bibfnamefont {T.}~\bibnamefont {Ishida}}, \bibinfo {author}
  {\bibfnamefont {H.}~\bibnamefont {Yoshida}}, \bibinfo {author} {\bibfnamefont
  {Y.}~\bibnamefont {Kawaguchi}}, \bibinfo {author} {\bibfnamefont
  {Y.}~\bibnamefont {Tanaka}}, \bibinfo {author} {\bibfnamefont
  {K.}~\bibnamefont {Saitoh}}, \bibinfo {author} {\bibfnamefont
  {N.}~\bibnamefont {Ikarashi}}, \bibinfo {author} {\bibfnamefont
  {M.}~\bibnamefont {Kuwahara}}, \ and\ \bibinfo {author} {\bibfnamefont
  {M.}~\bibnamefont {Nagao}},\ }\href {\doibase 10.1103/PhysRevLett.123.137203}
  {\bibfield  {journal} {\bibinfo  {journal} {Phys. Rev. Lett.}\ }\textbf
  {\bibinfo {volume} {123}},\ \bibinfo {pages} {137203} (\bibinfo {year}
  {2019})}\BibitemShut {NoStop}%
\bibitem [{\citenamefont {Romming}\ \emph {et~al.}(2013)\citenamefont
  {Romming}, \citenamefont {Hanneken}, \citenamefont {Menzel}, \citenamefont
  {Bickel}, \citenamefont {Wolter}, \citenamefont {von Bergmann}, \citenamefont
  {Kubetzka},\ and\ \citenamefont {Wiesendanger}}]{Romming2013b}%
  \BibitemOpen
  \bibfield  {author} {\bibinfo {author} {\bibfnamefont {N.}~\bibnamefont
  {Romming}}, \bibinfo {author} {\bibfnamefont {C.}~\bibnamefont {Hanneken}},
  \bibinfo {author} {\bibfnamefont {M.}~\bibnamefont {Menzel}}, \bibinfo
  {author} {\bibfnamefont {J.~E.}\ \bibnamefont {Bickel}}, \bibinfo {author}
  {\bibfnamefont {B.}~\bibnamefont {Wolter}}, \bibinfo {author} {\bibfnamefont
  {K.}~\bibnamefont {von Bergmann}}, \bibinfo {author} {\bibfnamefont
  {A.}~\bibnamefont {Kubetzka}}, \ and\ \bibinfo {author} {\bibfnamefont
  {R.}~\bibnamefont {Wiesendanger}},\ }\href {\doibase 10.1126/science.1240573}
  {\bibfield  {journal} {\bibinfo  {journal} {Science}\ }\textbf {\bibinfo
  {volume} {341}},\ \bibinfo {pages} {636} (\bibinfo {year}
  {2013})}\BibitemShut {NoStop}%
\bibitem [{\citenamefont {Yu}\ \emph {et~al.}(2012)\citenamefont {Yu},
  \citenamefont {Kanazawa}, \citenamefont {Zhang}, \citenamefont {Nagai},
  \citenamefont {Hara}, \citenamefont {Kimoto}, \citenamefont {Matsui},
  \citenamefont {Onose},\ and\ \citenamefont {Tokura}}]{Yu2012}%
  \BibitemOpen
  \bibfield  {author} {\bibinfo {author} {\bibfnamefont {X.}~\bibnamefont
  {Yu}}, \bibinfo {author} {\bibfnamefont {N.}~\bibnamefont {Kanazawa}},
  \bibinfo {author} {\bibfnamefont {W.}~\bibnamefont {Zhang}}, \bibinfo
  {author} {\bibfnamefont {T.}~\bibnamefont {Nagai}}, \bibinfo {author}
  {\bibfnamefont {T.}~\bibnamefont {Hara}}, \bibinfo {author} {\bibfnamefont
  {K.}~\bibnamefont {Kimoto}}, \bibinfo {author} {\bibfnamefont
  {Y.}~\bibnamefont {Matsui}}, \bibinfo {author} {\bibfnamefont
  {Y.}~\bibnamefont {Onose}}, \ and\ \bibinfo {author} {\bibfnamefont
  {Y.}~\bibnamefont {Tokura}},\ }\href {https://doi.org/10.1038/ncomms1990}
  {\bibfield  {journal} {\bibinfo  {journal} {Nature communications}\ }\textbf
  {\bibinfo {volume} {3}},\ \bibinfo {pages} {1} (\bibinfo {year}
  {2012})}\BibitemShut {NoStop}%
\bibitem [{\citenamefont {Zhao}\ \emph {et~al.}(2016)\citenamefont {Zhao},
  \citenamefont {Jin}, \citenamefont {Wang}, \citenamefont {Du}, \citenamefont
  {Zang}, \citenamefont {Tian}, \citenamefont {Che},\ and\ \citenamefont
  {Zhang}}]{Zhao2016}%
  \BibitemOpen
  \bibfield  {author} {\bibinfo {author} {\bibfnamefont {X.}~\bibnamefont
  {Zhao}}, \bibinfo {author} {\bibfnamefont {C.}~\bibnamefont {Jin}}, \bibinfo
  {author} {\bibfnamefont {C.}~\bibnamefont {Wang}}, \bibinfo {author}
  {\bibfnamefont {H.}~\bibnamefont {Du}}, \bibinfo {author} {\bibfnamefont
  {J.}~\bibnamefont {Zang}}, \bibinfo {author} {\bibfnamefont {M.}~\bibnamefont
  {Tian}}, \bibinfo {author} {\bibfnamefont {R.}~\bibnamefont {Che}}, \ and\
  \bibinfo {author} {\bibfnamefont {Y.}~\bibnamefont {Zhang}},\ }\href
  {\doibase 10.1073/pnas.1600197113} {\bibfield  {journal} {\bibinfo  {journal}
  {Proc. Natl. Acad. Sci. U. S. A.}\ }\textbf {\bibinfo {volume} {113}},\
  \bibinfo {pages} {4918} (\bibinfo {year} {2016})}\BibitemShut {NoStop}%
\bibitem [{\citenamefont {Nayak}\ \emph {et~al.}(2017)\citenamefont {Nayak},
  \citenamefont {Kumar}, \citenamefont {Ma}, \citenamefont {Werner},
  \citenamefont {Pippel}, \citenamefont {Sahoo}, \citenamefont {Damay},
  \citenamefont {R{\"{o}}{\ss}ler}, \citenamefont {Felser},\ and\ \citenamefont
  {Parkin}}]{Nayak2017}%
  \BibitemOpen
  \bibfield  {author} {\bibinfo {author} {\bibfnamefont {A.~K.}\ \bibnamefont
  {Nayak}}, \bibinfo {author} {\bibfnamefont {V.}~\bibnamefont {Kumar}},
  \bibinfo {author} {\bibfnamefont {T.}~\bibnamefont {Ma}}, \bibinfo {author}
  {\bibfnamefont {P.}~\bibnamefont {Werner}}, \bibinfo {author} {\bibfnamefont
  {E.}~\bibnamefont {Pippel}}, \bibinfo {author} {\bibfnamefont
  {R.}~\bibnamefont {Sahoo}}, \bibinfo {author} {\bibfnamefont
  {F.}~\bibnamefont {Damay}}, \bibinfo {author} {\bibfnamefont {U.~K.}\
  \bibnamefont {R{\"{o}}{\ss}ler}}, \bibinfo {author} {\bibfnamefont
  {C.}~\bibnamefont {Felser}}, \ and\ \bibinfo {author} {\bibfnamefont
  {S.~S.~P.}\ \bibnamefont {Parkin}},\ }\href {\doibase 10.1038/nature23466}
  {\bibfield  {journal} {\bibinfo  {journal} {Nature}\ }\textbf {\bibinfo
  {volume} {548}},\ \bibinfo {pages} {561} (\bibinfo {year}
  {2017})}\BibitemShut {NoStop}%
\bibitem [{\citenamefont {Vir}\ \emph {et~al.}(2019)\citenamefont {Vir},
  \citenamefont {Kumar}, \citenamefont {Borrmann}, \citenamefont
  {Jamijansuren}, \citenamefont {Kreiner}, \citenamefont {Shekhar},\ and\
  \citenamefont {Felser}}]{Vir2019}%
  \BibitemOpen
  \bibfield  {author} {\bibinfo {author} {\bibfnamefont {P.}~\bibnamefont
  {Vir}}, \bibinfo {author} {\bibfnamefont {N.}~\bibnamefont {Kumar}}, \bibinfo
  {author} {\bibfnamefont {H.}~\bibnamefont {Borrmann}}, \bibinfo {author}
  {\bibfnamefont {B.}~\bibnamefont {Jamijansuren}}, \bibinfo {author}
  {\bibfnamefont {G.}~\bibnamefont {Kreiner}}, \bibinfo {author} {\bibfnamefont
  {C.}~\bibnamefont {Shekhar}}, \ and\ \bibinfo {author} {\bibfnamefont
  {C.}~\bibnamefont {Felser}},\ }\href {\doibase 10.1021/acs.chemmater.9b02013}
  {\bibfield  {journal} {\bibinfo  {journal} {Chemistry of Materials}\ }\textbf
  {\bibinfo {volume} {31}},\ \bibinfo {pages} {5876} (\bibinfo {year}
  {2019})}\BibitemShut {NoStop}%
\bibitem [{\citenamefont {Jena}\ \emph {et~al.}(2020)\citenamefont {Jena},
  \citenamefont {Stinshoff}, \citenamefont {Saha}, \citenamefont {Srivastava},
  \citenamefont {Ma}, \citenamefont {Deniz}, \citenamefont {Werner},
  \citenamefont {Felser},\ and\ \citenamefont {Parkin}}]{jena2020observation}%
  \BibitemOpen
  \bibfield  {author} {\bibinfo {author} {\bibfnamefont {J.}~\bibnamefont
  {Jena}}, \bibinfo {author} {\bibfnamefont {R.}~\bibnamefont {Stinshoff}},
  \bibinfo {author} {\bibfnamefont {R.}~\bibnamefont {Saha}}, \bibinfo {author}
  {\bibfnamefont {A.~K.}\ \bibnamefont {Srivastava}}, \bibinfo {author}
  {\bibfnamefont {T.}~\bibnamefont {Ma}}, \bibinfo {author} {\bibfnamefont
  {H.}~\bibnamefont {Deniz}}, \bibinfo {author} {\bibfnamefont
  {P.}~\bibnamefont {Werner}}, \bibinfo {author} {\bibfnamefont
  {C.}~\bibnamefont {Felser}}, \ and\ \bibinfo {author} {\bibfnamefont
  {S.~S.~P.}\ \bibnamefont {Parkin}},\ }\href {\doibase
  10.1021/acs.nanolett.9b02973} {\bibfield  {journal} {\bibinfo  {journal}
  {Nano Letters}\ }\textbf {\bibinfo {volume} {20}},\ \bibinfo {pages} {59}
  (\bibinfo {year} {2020})}\BibitemShut {NoStop}%
\bibitem [{\citenamefont {Hayami}\ and\ \citenamefont
  {Motome}(2018)}]{Hayami2018}%
  \BibitemOpen
  \bibfield  {author} {\bibinfo {author} {\bibfnamefont {S.}~\bibnamefont
  {Hayami}}\ and\ \bibinfo {author} {\bibfnamefont {Y.}~\bibnamefont
  {Motome}},\ }\href {\doibase 10.1103/PhysRevLett.121.137202} {\bibfield
  {journal} {\bibinfo  {journal} {Phys. Rev. Lett.}\ }\textbf {\bibinfo
  {volume} {121}},\ \bibinfo {pages} {137202} (\bibinfo {year}
  {2018})}\BibitemShut {NoStop}%
\bibitem [{\citenamefont {Hayami}\ and\ \citenamefont
  {Motome}(2019)}]{Hayami2019}%
  \BibitemOpen
  \bibfield  {author} {\bibinfo {author} {\bibfnamefont {S.}~\bibnamefont
  {Hayami}}\ and\ \bibinfo {author} {\bibfnamefont {Y.}~\bibnamefont
  {Motome}},\ }\href {\doibase 10.1103/PhysRevB.99.094420} {\bibfield
  {journal} {\bibinfo  {journal} {Phys. Rev. B}\ }\textbf {\bibinfo {volume}
  {99}},\ \bibinfo {pages} {094420} (\bibinfo {year} {2019})}\BibitemShut
  {NoStop}%
\bibitem [{\citenamefont {Wang}\ \emph {et~al.}(2020)\citenamefont {Wang},
  \citenamefont {Su}, \citenamefont {Lin},\ and\ \citenamefont
  {Batista}}]{Wang2020}%
  \BibitemOpen
  \bibfield  {author} {\bibinfo {author} {\bibfnamefont {Z.}~\bibnamefont
  {Wang}}, \bibinfo {author} {\bibfnamefont {Y.}~\bibnamefont {Su}}, \bibinfo
  {author} {\bibfnamefont {S.-Z.}\ \bibnamefont {Lin}}, \ and\ \bibinfo
  {author} {\bibfnamefont {C.~D.}\ \bibnamefont {Batista}},\ }\href {\doibase
  10.1103/PhysRevLett.124.207201} {\bibfield  {journal} {\bibinfo  {journal}
  {Phys. Rev. Lett.}\ }\textbf {\bibinfo {volume} {124}},\ \bibinfo {pages}
  {207201} (\bibinfo {year} {2020})}\BibitemShut {NoStop}%
\bibitem [{\citenamefont {Bezvershenko}\ \emph {et~al.}(2018)\citenamefont
  {Bezvershenko}, \citenamefont {Kolezhuk},\ and\ \citenamefont
  {Ivanov}}]{Bezvershenko2018}%
  \BibitemOpen
  \bibfield  {author} {\bibinfo {author} {\bibfnamefont {A.~V.}\ \bibnamefont
  {Bezvershenko}}, \bibinfo {author} {\bibfnamefont {A.~K.}\ \bibnamefont
  {Kolezhuk}}, \ and\ \bibinfo {author} {\bibfnamefont {B.~A.}\ \bibnamefont
  {Ivanov}},\ }\href {\doibase 10.1103/PhysRevB.97.054408} {\bibfield
  {journal} {\bibinfo  {journal} {Phys. Rev. B}\ }\textbf {\bibinfo {volume}
  {97}},\ \bibinfo {pages} {054408} (\bibinfo {year} {2018})}\BibitemShut
  {NoStop}%
\bibitem [{\citenamefont {Hayami}\ and\ \citenamefont
  {Yambe}(2021)}]{hayami2021field}%
  \BibitemOpen
  \bibfield  {author} {\bibinfo {author} {\bibfnamefont {S.}~\bibnamefont
  {Hayami}}\ and\ \bibinfo {author} {\bibfnamefont {R.}~\bibnamefont {Yambe}},\
  }\href {https://journals.jps.jp/doi/full/10.7566/JPSJ.90.073705} {\bibfield
  {journal} {\bibinfo  {journal} {Journal of the Physical Society of Japan}\
  }\textbf {\bibinfo {volume} {90}},\ \bibinfo {pages} {073705} (\bibinfo
  {year} {2021})}\BibitemShut {NoStop}%
\bibitem [{\citenamefont {Jin}\ \emph {et~al.}(2021)\citenamefont {Jin},
  \citenamefont {Xi}, \citenamefont {Mei},\ and\ \citenamefont
  {Liu}}]{jin2021field}%
  \BibitemOpen
  \bibfield  {author} {\bibinfo {author} {\bibfnamefont {L.}~\bibnamefont
  {Jin}}, \bibinfo {author} {\bibfnamefont {B.}~\bibnamefont {Xi}}, \bibinfo
  {author} {\bibfnamefont {J.-W.}\ \bibnamefont {Mei}}, \ and\ \bibinfo
  {author} {\bibfnamefont {Y.}~\bibnamefont {Liu}},\ }\href
  {https://arxiv.org/abs/2106.00207} {\bibfield  {journal} {\bibinfo  {journal}
  {arXiv preprint arXiv:2106.00207}\ } (\bibinfo {year} {2021})}\BibitemShut
  {NoStop}%
\bibitem [{\citenamefont {Chen}\ \emph {et~al.}(2016)\citenamefont {Chen},
  \citenamefont {Zhang},\ and\ \citenamefont {Liu}}]{chen2016}%
  \BibitemOpen
  \bibfield  {author} {\bibinfo {author} {\bibfnamefont {J.~P.}\ \bibnamefont
  {Chen}}, \bibinfo {author} {\bibfnamefont {D.-W.}\ \bibnamefont {Zhang}}, \
  and\ \bibinfo {author} {\bibfnamefont {J.~M.}\ \bibnamefont {Liu}},\ }\href
  {\doibase 10.1038/srep29126} {\bibfield  {journal} {\bibinfo  {journal} {Sci.
  Rep.}\ }\textbf {\bibinfo {volume} {6}},\ \bibinfo {pages} {29126} (\bibinfo
  {year} {2016})}\BibitemShut {NoStop}%
\bibitem [{\citenamefont {Roessler}\ \emph {et~al.}(2006)\citenamefont
  {Roessler}, \citenamefont {Bogdanov},\ and\ \citenamefont
  {Pfleiderer}}]{roessler}%
  \BibitemOpen
  \bibfield  {author} {\bibinfo {author} {\bibfnamefont {U.~K.}\ \bibnamefont
  {Roessler}}, \bibinfo {author} {\bibfnamefont {A.}~\bibnamefont {Bogdanov}},
  \ and\ \bibinfo {author} {\bibfnamefont {C.}~\bibnamefont {Pfleiderer}},\
  }\href {https://doi.org/10.1038/nature05056} {\bibfield  {journal} {\bibinfo
  {journal} {Nature}\ }\textbf {\bibinfo {volume} {442}},\ \bibinfo {pages}
  {797} (\bibinfo {year} {2006})}\BibitemShut {NoStop}%
\bibitem [{\citenamefont {Mohanta}\ \emph {et~al.}(2019)\citenamefont
  {Mohanta}, \citenamefont {Dagotto},\ and\ \citenamefont
  {Okamoto}}]{mohanta2019}%
  \BibitemOpen
  \bibfield  {author} {\bibinfo {author} {\bibfnamefont {N.}~\bibnamefont
  {Mohanta}}, \bibinfo {author} {\bibfnamefont {E.}~\bibnamefont {Dagotto}}, \
  and\ \bibinfo {author} {\bibfnamefont {S.}~\bibnamefont {Okamoto}},\ }\href
  {https://journals.aps.org/prb/abstract/10.1103/PhysRevB.100.064429}
  {\bibfield  {journal} {\bibinfo  {journal} {Physical Review B}\ }\textbf
  {\bibinfo {volume} {100}},\ \bibinfo {pages} {064429} (\bibinfo {year}
  {2019})}\BibitemShut {NoStop}%
\bibitem [{\citenamefont {Iwasaki}\ \emph {et~al.}(2014)\citenamefont
  {Iwasaki}, \citenamefont {Beekman},\ and\ \citenamefont
  {Nagaosa}}]{iwasaki2014}%
  \BibitemOpen
  \bibfield  {author} {\bibinfo {author} {\bibfnamefont {J.}~\bibnamefont
  {Iwasaki}}, \bibinfo {author} {\bibfnamefont {A.~J.}\ \bibnamefont
  {Beekman}}, \ and\ \bibinfo {author} {\bibfnamefont {N.}~\bibnamefont
  {Nagaosa}},\ }\href
  {https://journals.aps.org/prb/abstract/10.1103/PhysRevB.89.064412} {\bibfield
   {journal} {\bibinfo  {journal} {Physical Review B}\ }\textbf {\bibinfo
  {volume} {89}},\ \bibinfo {pages} {064412} (\bibinfo {year}
  {2014})}\BibitemShut {NoStop}%
\bibitem [{\citenamefont {Yi}\ \emph {et~al.}(2009)\citenamefont {Yi},
  \citenamefont {Onoda}, \citenamefont {Nagaosa},\ and\ \citenamefont
  {Han}}]{Yi2009}%
  \BibitemOpen
  \bibfield  {author} {\bibinfo {author} {\bibfnamefont {S.~D.}\ \bibnamefont
  {Yi}}, \bibinfo {author} {\bibfnamefont {S.}~\bibnamefont {Onoda}}, \bibinfo
  {author} {\bibfnamefont {N.}~\bibnamefont {Nagaosa}}, \ and\ \bibinfo
  {author} {\bibfnamefont {J.~H.}\ \bibnamefont {Han}},\ }\href {\doibase
  10.1103/PhysRevB.80.054416} {\bibfield  {journal} {\bibinfo  {journal} {Phys.
  Rev. B}\ }\textbf {\bibinfo {volume} {80}},\ \bibinfo {pages} {054416}
  (\bibinfo {year} {2009})}\BibitemShut {NoStop}%
\bibitem [{\citenamefont {Hoffmann}\ \emph {et~al.}(2017)\citenamefont
  {Hoffmann}, \citenamefont {Zimmermann}, \citenamefont {M{\"u}ller},
  \citenamefont {Sch{\"u}rhoff}, \citenamefont {Kiselev}, \citenamefont
  {Melcher},\ and\ \citenamefont {Bl{\"u}gel}}]{hoffmann2017antiskyrmions}%
  \BibitemOpen
  \bibfield  {author} {\bibinfo {author} {\bibfnamefont {M.}~\bibnamefont
  {Hoffmann}}, \bibinfo {author} {\bibfnamefont {B.}~\bibnamefont
  {Zimmermann}}, \bibinfo {author} {\bibfnamefont {G.~P.}\ \bibnamefont
  {M{\"u}ller}}, \bibinfo {author} {\bibfnamefont {D.}~\bibnamefont
  {Sch{\"u}rhoff}}, \bibinfo {author} {\bibfnamefont {N.~S.}\ \bibnamefont
  {Kiselev}}, \bibinfo {author} {\bibfnamefont {C.}~\bibnamefont {Melcher}}, \
  and\ \bibinfo {author} {\bibfnamefont {S.}~\bibnamefont {Bl{\"u}gel}},\
  }\href {https://doi.org/10.1038/s41467-017-00313-0} {\bibfield  {journal}
  {\bibinfo  {journal} {Nature communications}\ }\textbf {\bibinfo {volume}
  {8}},\ \bibinfo {pages} {1} (\bibinfo {year} {2017})}\BibitemShut {NoStop}%
\bibitem [{\citenamefont {Huang}\ \emph {et~al.}(2017)\citenamefont {Huang},
  \citenamefont {Zhou}, \citenamefont {Chen}, \citenamefont {Shen},
  \citenamefont {Schmid}, \citenamefont {Liu},\ and\ \citenamefont
  {Wu}}]{huang2017stabilization}%
  \BibitemOpen
  \bibfield  {author} {\bibinfo {author} {\bibfnamefont {S.}~\bibnamefont
  {Huang}}, \bibinfo {author} {\bibfnamefont {C.}~\bibnamefont {Zhou}},
  \bibinfo {author} {\bibfnamefont {G.}~\bibnamefont {Chen}}, \bibinfo {author}
  {\bibfnamefont {H.}~\bibnamefont {Shen}}, \bibinfo {author} {\bibfnamefont
  {A.~K.}\ \bibnamefont {Schmid}}, \bibinfo {author} {\bibfnamefont
  {K.}~\bibnamefont {Liu}}, \ and\ \bibinfo {author} {\bibfnamefont
  {Y.}~\bibnamefont {Wu}},\ }\href
  {https://journals.aps.org/prb/abstract/10.1103/PhysRevB.96.144412} {\bibfield
   {journal} {\bibinfo  {journal} {Physical Review B}\ }\textbf {\bibinfo
  {volume} {96}},\ \bibinfo {pages} {144412} (\bibinfo {year}
  {2017})}\BibitemShut {NoStop}%
\bibitem [{\citenamefont {Qiu}\ \emph {et~al.}(2020)\citenamefont {Qiu},
  \citenamefont {Xia}, \citenamefont {Feng}, \citenamefont {Shen},
  \citenamefont {Morvan}, \citenamefont {Zhang}, \citenamefont {Liu},
  \citenamefont {Xie}, \citenamefont {Zhou},\ and\ \citenamefont
  {Zhao}}]{qiu2020dynamics}%
  \BibitemOpen
  \bibfield  {author} {\bibinfo {author} {\bibfnamefont {L.}~\bibnamefont
  {Qiu}}, \bibinfo {author} {\bibfnamefont {J.}~\bibnamefont {Xia}}, \bibinfo
  {author} {\bibfnamefont {Y.}~\bibnamefont {Feng}}, \bibinfo {author}
  {\bibfnamefont {L.}~\bibnamefont {Shen}}, \bibinfo {author} {\bibfnamefont
  {F.~J.}\ \bibnamefont {Morvan}}, \bibinfo {author} {\bibfnamefont
  {X.}~\bibnamefont {Zhang}}, \bibinfo {author} {\bibfnamefont
  {X.}~\bibnamefont {Liu}}, \bibinfo {author} {\bibfnamefont {L.}~\bibnamefont
  {Xie}}, \bibinfo {author} {\bibfnamefont {Y.}~\bibnamefont {Zhou}}, \ and\
  \bibinfo {author} {\bibfnamefont {G.}~\bibnamefont {Zhao}},\ }\href
  {https://www.sciencedirect.com/science/article/pii/S0304885319302562}
  {\bibfield  {journal} {\bibinfo  {journal} {Journal of Magnetism and Magnetic
  Materials}\ }\textbf {\bibinfo {volume} {496}},\ \bibinfo {pages} {165922}
  (\bibinfo {year} {2020})}\BibitemShut {NoStop}%
\bibitem [{\citenamefont {Koshibae}\ and\ \citenamefont
  {Nagaosa}(2016)}]{koshibae2016theory}%
  \BibitemOpen
  \bibfield  {author} {\bibinfo {author} {\bibfnamefont {W.}~\bibnamefont
  {Koshibae}}\ and\ \bibinfo {author} {\bibfnamefont {N.}~\bibnamefont
  {Nagaosa}},\ }\href {https://doi.org/10.1038/ncomms10542} {\bibfield
  {journal} {\bibinfo  {journal} {Nature communications}\ }\textbf {\bibinfo
  {volume} {7}},\ \bibinfo {pages} {1} (\bibinfo {year} {2016})}\BibitemShut
  {NoStop}%
\bibitem [{\citenamefont {Kathyat}\ \emph {et~al.}(2020)\citenamefont
  {Kathyat}, \citenamefont {Mukherjee},\ and\ \citenamefont
  {Kumar}}]{Kathyat2020a}%
  \BibitemOpen
  \bibfield  {author} {\bibinfo {author} {\bibfnamefont {D.~S.}\ \bibnamefont
  {Kathyat}}, \bibinfo {author} {\bibfnamefont {A.}~\bibnamefont {Mukherjee}},
  \ and\ \bibinfo {author} {\bibfnamefont {S.}~\bibnamefont {Kumar}},\ }\href
  {\doibase 10.1103/PhysRevB.102.075106} {\bibfield  {journal} {\bibinfo
  {journal} {Phys. Rev. B}\ }\textbf {\bibinfo {volume} {102}},\ \bibinfo
  {pages} {075106} (\bibinfo {year} {2020})}\BibitemShut {NoStop}%
\bibitem [{\citenamefont {Kathyat}\ \emph {et~al.}(2021)\citenamefont
  {Kathyat}, \citenamefont {Mukherjee},\ and\ \citenamefont
  {Kumar}}]{Kathyat2021}%
  \BibitemOpen
  \bibfield  {author} {\bibinfo {author} {\bibfnamefont {D.~S.}\ \bibnamefont
  {Kathyat}}, \bibinfo {author} {\bibfnamefont {A.}~\bibnamefont {Mukherjee}},
  \ and\ \bibinfo {author} {\bibfnamefont {S.}~\bibnamefont {Kumar}},\ }\href
  {\doibase 10.1103/PhysRevB.103.035111} {\bibfield  {journal} {\bibinfo
  {journal} {Phys. Rev. B}\ }\textbf {\bibinfo {volume} {103}},\ \bibinfo
  {pages} {035111} (\bibinfo {year} {2021})}\BibitemShut {NoStop}%
\bibitem [{\citenamefont {Okubo}\ \emph {et~al.}(2012)\citenamefont {Okubo},
  \citenamefont {Chung},\ and\ \citenamefont
  {Kawamura}}]{PhysRevLett.108.017206}%
  \BibitemOpen
  \bibfield  {author} {\bibinfo {author} {\bibfnamefont {T.}~\bibnamefont
  {Okubo}}, \bibinfo {author} {\bibfnamefont {S.}~\bibnamefont {Chung}}, \ and\
  \bibinfo {author} {\bibfnamefont {H.}~\bibnamefont {Kawamura}},\ }\href
  {\doibase 10.1103/PhysRevLett.108.017206} {\bibfield  {journal} {\bibinfo
  {journal} {Phys. Rev. Lett.}\ }\textbf {\bibinfo {volume} {108}},\ \bibinfo
  {pages} {017206} (\bibinfo {year} {2012})}\BibitemShut {NoStop}%
\bibitem [{\citenamefont {Araki}(2020)}]{araki2020magnetic}%
  \BibitemOpen
  \bibfield  {author} {\bibinfo {author} {\bibfnamefont {Y.}~\bibnamefont
  {Araki}},\ }\href
  {https://onlinelibrary.wiley.com/doi/10.1002/andp.201900287} {\bibfield
  {journal} {\bibinfo  {journal} {Annalen der Physik}\ }\textbf {\bibinfo
  {volume} {532}},\ \bibinfo {pages} {1900287} (\bibinfo {year}
  {2020})}\BibitemShut {NoStop}%
\bibitem [{\citenamefont {Wu}\ \emph {et~al.}(2020)\citenamefont {Wu},
  \citenamefont {Sun}, \citenamefont {Hsieh}, \citenamefont {Chen},
  \citenamefont {Kakarla}, \citenamefont {Deng}, \citenamefont {Chu},\ and\
  \citenamefont {Yang}}]{wu2020observation}%
  \BibitemOpen
  \bibfield  {author} {\bibinfo {author} {\bibfnamefont {H.}~\bibnamefont
  {Wu}}, \bibinfo {author} {\bibfnamefont {P.}~\bibnamefont {Sun}}, \bibinfo
  {author} {\bibfnamefont {D.}~\bibnamefont {Hsieh}}, \bibinfo {author}
  {\bibfnamefont {H.}~\bibnamefont {Chen}}, \bibinfo {author} {\bibfnamefont
  {D.~C.}\ \bibnamefont {Kakarla}}, \bibinfo {author} {\bibfnamefont
  {L.}~\bibnamefont {Deng}}, \bibinfo {author} {\bibfnamefont {C.-W.}\
  \bibnamefont {Chu}}, \ and\ \bibinfo {author} {\bibfnamefont
  {H.}~\bibnamefont {Yang}},\ }\href
  {https://www.sciencedirect.com/science/article/pii/S2542529320300134}
  {\bibfield  {journal} {\bibinfo  {journal} {Materials Today Physics}\
  }\textbf {\bibinfo {volume} {12}},\ \bibinfo {pages} {100189} (\bibinfo
  {year} {2020})}\BibitemShut {NoStop}%
\bibitem [{\citenamefont {Puphal}\ \emph {et~al.}(2020)\citenamefont {Puphal},
  \citenamefont {Pomjakushin}, \citenamefont {Kanazawa}, \citenamefont
  {Ukleev}, \citenamefont {Gawryluk}, \citenamefont {Ma}, \citenamefont
  {Naamneh}, \citenamefont {Plumb}, \citenamefont {Keller}, \citenamefont
  {Cubitt}, \citenamefont {Pomjakushina},\ and\ \citenamefont
  {White}}]{PhysRevLett.124.017202}%
  \BibitemOpen
  \bibfield  {author} {\bibinfo {author} {\bibfnamefont {P.}~\bibnamefont
  {Puphal}}, \bibinfo {author} {\bibfnamefont {V.}~\bibnamefont {Pomjakushin}},
  \bibinfo {author} {\bibfnamefont {N.}~\bibnamefont {Kanazawa}}, \bibinfo
  {author} {\bibfnamefont {V.}~\bibnamefont {Ukleev}}, \bibinfo {author}
  {\bibfnamefont {D.~J.}\ \bibnamefont {Gawryluk}}, \bibinfo {author}
  {\bibfnamefont {J.}~\bibnamefont {Ma}}, \bibinfo {author} {\bibfnamefont
  {M.}~\bibnamefont {Naamneh}}, \bibinfo {author} {\bibfnamefont {N.~C.}\
  \bibnamefont {Plumb}}, \bibinfo {author} {\bibfnamefont {L.}~\bibnamefont
  {Keller}}, \bibinfo {author} {\bibfnamefont {R.}~\bibnamefont {Cubitt}},
  \bibinfo {author} {\bibfnamefont {E.}~\bibnamefont {Pomjakushina}}, \ and\
  \bibinfo {author} {\bibfnamefont {J.~S.}\ \bibnamefont {White}},\ }\href
  {\doibase 10.1103/PhysRevLett.124.017202} {\bibfield  {journal} {\bibinfo
  {journal} {Phys. Rev. Lett.}\ }\textbf {\bibinfo {volume} {124}},\ \bibinfo
  {pages} {017202} (\bibinfo {year} {2020})}\BibitemShut {NoStop}%
\bibitem [{\citenamefont {Redies}\ \emph {et~al.}(2020)\citenamefont {Redies},
  \citenamefont {Lux}, \citenamefont {Hanke}, \citenamefont {Buhl},
  \citenamefont {Bl\"ugel},\ and\ \citenamefont
  {Mokrousov}}]{PhysRevB.102.184407}%
  \BibitemOpen
  \bibfield  {author} {\bibinfo {author} {\bibfnamefont {M.}~\bibnamefont
  {Redies}}, \bibinfo {author} {\bibfnamefont {F.~R.}\ \bibnamefont {Lux}},
  \bibinfo {author} {\bibfnamefont {J.-P.}\ \bibnamefont {Hanke}}, \bibinfo
  {author} {\bibfnamefont {P.~M.}\ \bibnamefont {Buhl}}, \bibinfo {author}
  {\bibfnamefont {S.}~\bibnamefont {Bl\"ugel}}, \ and\ \bibinfo {author}
  {\bibfnamefont {Y.}~\bibnamefont {Mokrousov}},\ }\href {\doibase
  10.1103/PhysRevB.102.184407} {\bibfield  {journal} {\bibinfo  {journal}
  {Phys. Rev. B}\ }\textbf {\bibinfo {volume} {102}},\ \bibinfo {pages}
  {184407} (\bibinfo {year} {2020})}\BibitemShut {NoStop}%
\bibitem [{\citenamefont {Shanavas}\ and\ \citenamefont
  {Satpathy}(2016)}]{PhysRevB.93.195101}%
  \BibitemOpen
  \bibfield  {author} {\bibinfo {author} {\bibfnamefont {K.~V.}\ \bibnamefont
  {Shanavas}}\ and\ \bibinfo {author} {\bibfnamefont {S.}~\bibnamefont
  {Satpathy}},\ }\href {\doibase 10.1103/PhysRevB.93.195101} {\bibfield
  {journal} {\bibinfo  {journal} {Phys. Rev. B}\ }\textbf {\bibinfo {volume}
  {93}},\ \bibinfo {pages} {195101} (\bibinfo {year} {2016})}\BibitemShut
  {NoStop}%
\bibitem [{\citenamefont {Choi}\ \emph {et~al.}(2019)\citenamefont {Choi},
  \citenamefont {Tai},\ and\ \citenamefont {Zhu}}]{PhysRevB.99.134437}%
  \BibitemOpen
  \bibfield  {author} {\bibinfo {author} {\bibfnamefont {H.}~\bibnamefont
  {Choi}}, \bibinfo {author} {\bibfnamefont {Y.-Y.}\ \bibnamefont {Tai}}, \
  and\ \bibinfo {author} {\bibfnamefont {J.-X.}\ \bibnamefont {Zhu}},\ }\href
  {\doibase 10.1103/PhysRevB.99.134437} {\bibfield  {journal} {\bibinfo
  {journal} {Phys. Rev. B}\ }\textbf {\bibinfo {volume} {99}},\ \bibinfo
  {pages} {134437} (\bibinfo {year} {2019})}\BibitemShut {NoStop}%
\bibitem [{\citenamefont {Nomoto}\ \emph {et~al.}(2020)\citenamefont {Nomoto},
  \citenamefont {Koretsune},\ and\ \citenamefont
  {Arita}}]{PhysRevLett.125.117204}%
  \BibitemOpen
  \bibfield  {author} {\bibinfo {author} {\bibfnamefont {T.}~\bibnamefont
  {Nomoto}}, \bibinfo {author} {\bibfnamefont {T.}~\bibnamefont {Koretsune}}, \
  and\ \bibinfo {author} {\bibfnamefont {R.}~\bibnamefont {Arita}},\ }\href
  {\doibase 10.1103/PhysRevLett.125.117204} {\bibfield  {journal} {\bibinfo
  {journal} {Phys. Rev. Lett.}\ }\textbf {\bibinfo {volume} {125}},\ \bibinfo
  {pages} {117204} (\bibinfo {year} {2020})}\BibitemShut {NoStop}%
\bibitem [{\citenamefont {Koretsune}\ \emph {et~al.}(2015)\citenamefont
  {Koretsune}, \citenamefont {Nagaosa},\ and\ \citenamefont
  {Arita}}]{koretsune2015control}%
  \BibitemOpen
  \bibfield  {author} {\bibinfo {author} {\bibfnamefont {T.}~\bibnamefont
  {Koretsune}}, \bibinfo {author} {\bibfnamefont {N.}~\bibnamefont {Nagaosa}},
  \ and\ \bibinfo {author} {\bibfnamefont {R.}~\bibnamefont {Arita}},\ }\href
  {https://doi.org/10.1038/srep13302} {\bibfield  {journal} {\bibinfo
  {journal} {Scientific reports}\ }\textbf {\bibinfo {volume} {5}},\ \bibinfo
  {pages} {1} (\bibinfo {year} {2015})}\BibitemShut {NoStop}%
\bibitem [{\citenamefont {Martin}\ and\ \citenamefont
  {Batista}(2008)}]{PhysRevLett.101.156402}%
  \BibitemOpen
  \bibfield  {author} {\bibinfo {author} {\bibfnamefont {I.}~\bibnamefont
  {Martin}}\ and\ \bibinfo {author} {\bibfnamefont {C.~D.}\ \bibnamefont
  {Batista}},\ }\href {\doibase 10.1103/PhysRevLett.101.156402} {\bibfield
  {journal} {\bibinfo  {journal} {Phys. Rev. Lett.}\ }\textbf {\bibinfo
  {volume} {101}},\ \bibinfo {pages} {156402} (\bibinfo {year}
  {2008})}\BibitemShut {NoStop}%
\bibitem [{\citenamefont {Pasrija}\ and\ \citenamefont
  {Kumar}(2016)}]{PhysRevB.93.195110}%
  \BibitemOpen
  \bibfield  {author} {\bibinfo {author} {\bibfnamefont {K.}~\bibnamefont
  {Pasrija}}\ and\ \bibinfo {author} {\bibfnamefont {S.}~\bibnamefont
  {Kumar}},\ }\href {\doibase 10.1103/PhysRevB.93.195110} {\bibfield  {journal}
  {\bibinfo  {journal} {Phys. Rev. B}\ }\textbf {\bibinfo {volume} {93}},\
  \bibinfo {pages} {195110} (\bibinfo {year} {2016})}\BibitemShut {NoStop}%
\bibitem [{\citenamefont {Kumar}\ and\ \citenamefont
  {Majumdar}(2006)}]{kumar2006}%
  \BibitemOpen
  \bibfield  {author} {\bibinfo {author} {\bibfnamefont {S.}~\bibnamefont
  {Kumar}}\ and\ \bibinfo {author} {\bibfnamefont {P.}~\bibnamefont
  {Majumdar}},\ }\href {https://doi.org/10.1140/epjb/e2006-00173-2} {\bibfield
  {journal} {\bibinfo  {journal} {The European Physical Journal B-Condensed
  Matter and Complex Systems}\ }\textbf {\bibinfo {volume} {50}},\ \bibinfo
  {pages} {571} (\bibinfo {year} {2006})}\BibitemShut {NoStop}%
\bibitem [{\citenamefont {Mukherjee}\ \emph {et~al.}(2015)\citenamefont
  {Mukherjee}, \citenamefont {Patel}, \citenamefont {Bishop},\ and\
  \citenamefont {Dagotto}}]{mukherjee2015}%
  \BibitemOpen
  \bibfield  {author} {\bibinfo {author} {\bibfnamefont {A.}~\bibnamefont
  {Mukherjee}}, \bibinfo {author} {\bibfnamefont {N.~D.}\ \bibnamefont
  {Patel}}, \bibinfo {author} {\bibfnamefont {C.}~\bibnamefont {Bishop}}, \
  and\ \bibinfo {author} {\bibfnamefont {E.}~\bibnamefont {Dagotto}},\ }\href
  {https://journals.aps.org/pre/abstract/10.1103/PhysRevE.91.063303} {\bibfield
   {journal} {\bibinfo  {journal} {Physical Review E}\ }\textbf {\bibinfo
  {volume} {91}},\ \bibinfo {pages} {063303} (\bibinfo {year}
  {2015})}\BibitemShut {NoStop}%
\bibitem [{\citenamefont {Xu}\ and\ \citenamefont
  {Xiang}(2015)}]{PhysRevB.92.121112}%
  \BibitemOpen
  \bibfield  {author} {\bibinfo {author} {\bibfnamefont {K.}~\bibnamefont
  {Xu}}\ and\ \bibinfo {author} {\bibfnamefont {H.~J.}\ \bibnamefont {Xiang}},\
  }\href {\doibase 10.1103/PhysRevB.92.121112} {\bibfield  {journal} {\bibinfo
  {journal} {Phys. Rev. B}\ }\textbf {\bibinfo {volume} {92}},\ \bibinfo
  {pages} {121112} (\bibinfo {year} {2015})}\BibitemShut {NoStop}%
\bibitem [{\citenamefont {Bhowal}\ and\ \citenamefont
  {Satpathy}(2019{\natexlab{a}})}]{bhowal2019electric}%
  \BibitemOpen
  \bibfield  {author} {\bibinfo {author} {\bibfnamefont {S.}~\bibnamefont
  {Bhowal}}\ and\ \bibinfo {author} {\bibfnamefont {S.}~\bibnamefont
  {Satpathy}},\ }\href {https://doi.org/10.1038/s41524-019-0198-8} {\bibfield
  {journal} {\bibinfo  {journal} {npj Computational Materials}\ }\textbf
  {\bibinfo {volume} {5}},\ \bibinfo {pages} {1} (\bibinfo {year}
  {2019}{\natexlab{a}})}\BibitemShut {NoStop}%
\bibitem [{\citenamefont {Bhowal}\ and\ \citenamefont
  {Satpathy}(2019{\natexlab{b}})}]{PhysRevB.99.245145}%
  \BibitemOpen
  \bibfield  {author} {\bibinfo {author} {\bibfnamefont {S.}~\bibnamefont
  {Bhowal}}\ and\ \bibinfo {author} {\bibfnamefont {S.}~\bibnamefont
  {Satpathy}},\ }\href {\doibase 10.1103/PhysRevB.99.245145} {\bibfield
  {journal} {\bibinfo  {journal} {Phys. Rev. B}\ }\textbf {\bibinfo {volume}
  {99}},\ \bibinfo {pages} {245145} (\bibinfo {year}
  {2019}{\natexlab{b}})}\BibitemShut {NoStop}%
\end{thebibliography}
%

\end{document}